\documentclass[conference]{IEEEtran}
\pdfoutput=1 
\usepackage{blindtext, graphicx}
\usepackage{amsmath}
\usepackage[table,xcdraw]{xcolor}

\begin{document}

\title{Solar Power Smoothing in a Nanogrid Testbed} 

	\author{\IEEEauthorblockN{Hossein Panamtash and Qun Zhou Sun
	}
		\IEEEauthorblockA{\textit{Department of Electrical and Computer Engineering} \\
			\textit{University of Central Florida}, 
			Orlando, FL, USA \\
			h.panamtash@knights.ucf.edu,
            qz.sun@ucf.edu }
            
    \and
            
    \IEEEauthorblockN{Rubin York, Paul Brooker and Justin Kramer  
	}
        \IEEEauthorblockA{\textit{Orlando Utilities Commission} 
			Orlando, FL, USA \\
			ryork@ouc.com,
            pbrooker@ouc.com,\\
            jkramer@ouc.com}
		
	}

	\IEEEoverridecommandlockouts
\IEEEpubid{\makebox[\columnwidth]{978-1-7281-8192-9/21/\$31.00~\copyright2021 IEEE \hfill}
\hspace{\columnsep}\makebox[\columnwidth]{ }}


\maketitle

\IEEEpubidadjcol
	
\maketitle

\begin{abstract}
High penetration of solar power introduces new challenges in the operation of distribution systems. Considering the highly volatile nature of solar power output due to changes in cloud coverage, maintaining the power balance and operating within ramp rate limits can be an issue. Great benefits can be brought to the grid by smoothing solar power output at individual sites equipped with flexible resources such as electrical vehicles and battery storage systems. This paper proposes several approaches to a solar smoothing application by utilizing battery storage and EV charging control in a "Nanogrid" testbed located at a utility in Florida. The control algorithms focus on both real-time application and predictive control depending on forecasts. The solar smoothing models are then compared using real data from the Nanogrid site to present the effectiveness of the proposed models and compare their results. Furthermore, the control methods are applied to the Orlando Utilities Commission (OUC) Nanogrid to confirm the simulation results.
\end{abstract}


\IEEEpeerreviewmaketitle
\section{Introduction}

The power grid is evolving to include an increasing penetration level of renewable energy \cite{8440483}. While a higher level of renewable generation results in less pollution and cheaper energy generation, it also introduces new challenges to the power system operation \cite{9183699}. The uncertainty associated with renewable energy including Photo-Voltaic (PV) generation and the very volatile nature of the power output makes it difficult for the power system operator to maintain the power balance in the system \cite{PANAMTASH2020336}. The ramping capability of the conventional generation units in the power system is limited therefore it may not be able to change the power output as fast and frequently as the changes in solar power output \cite{9072298}. The output of PV units can change drastically within a very short period of time which could be more than 50\% of the PV capacity within a minute \cite{9449746}.

Possible solutions to the ramping problem in the distribution grid include utilizing energy storage devices and mitigating the controllable load in the system \cite{6804417}. The energy storage devices can be employed to help the power balance when the ramping capability of the system is not sufficient to maintain the balance \cite{6763050}. The energy storage devices installed alongside the PV system will start charging when there is a sudden rise in solar generation and discharge when the solar power output drops significantly due to cloud coverage \cite{8370779}. While the storage system can help significantly with the power balance problem, its finite capacity limits the level of contribution to this problem. More importantly, if the energy storage system is discharged during the drops in solar power output and then charged during the rise, its energy level will decrease on account of the energy loss and efficiency of the storage device.

Load mitigation is the further solution to the power balance problem \cite{8445610,7999298}. If there are controllable loads connected to the distribution system which are capable of changing very fast, the system operator can also depend on lowering the system load during drops in solar power output alongside energy storage which in turn prevents the energy storage system from draining. One of the most appropriate loads in the distribution system for this task is the (Electrical Vehicle) EV charging load \cite{7264982}. The EV load may be changed substantially within a few moment's notice. Additionally, if the EV connection time is long enough, utilizing the EV charging rate to maintain the power balance will not decrease the amount of energy transferred to the EV.

Many efforts have been dedicated to solar power smoothing in the literature \cite{7858894,9364421}. While there are reliable control algorithms with acceptable results using battery storage for solar power smoothing \cite{8990151}, most of the studies have not considered the effects of EV charging and possible mitigation of the EV load within the system \cite{6777277}. Although addition of the EV charging to the control algorithm will introduce new constraints for charging the EVs, the additional source available to the system operator can play a crucial role in maintaining the ramp rate changes within the limits.

The challenge for controlling the battery system and EV load in the system is finding the optimal power level depending on the previous and current solar generation and load as well as the future changes \cite{9373906}. In this paper, we first generate solar power forecasts to predict the future changes in solar power output. Then the historical data and the forecasts are used to find the optimal power balance without violating the ramp rate limits. If the violation is unpreventable, then the objective would be to minimize the ramp rate violation. Two target curves are proposed for optimal PV smoothing in this paper. The results of the proposed methods are compared to determine the effectiveness in terms of ramp rate violation as well as other factors such as EV charging and battery State of Charge (SoC) level.

The control methods in this paper are implemented and compared in simulation. Additionally, the control methods are applied to the Nanogrid site at an Orlando Utilities Commission (OUC) location to confirm the results.

The remainder of the paper is organized as follows: Overview of the problem and the real-time solution, as well as proposed predictive solar smoothing, are explained in section \ref{Solar Smoothing}, the Nanogrid system used in this study and the results for the different control algorithms are presented and compared in section \ref{Case Study}, and section \ref{Conclusion} provides the concluding remarks.

\section{Solar Smoothing}
\label{Solar Smoothing}

\begin{figure}[h!]
\includegraphics[width=3.8in]{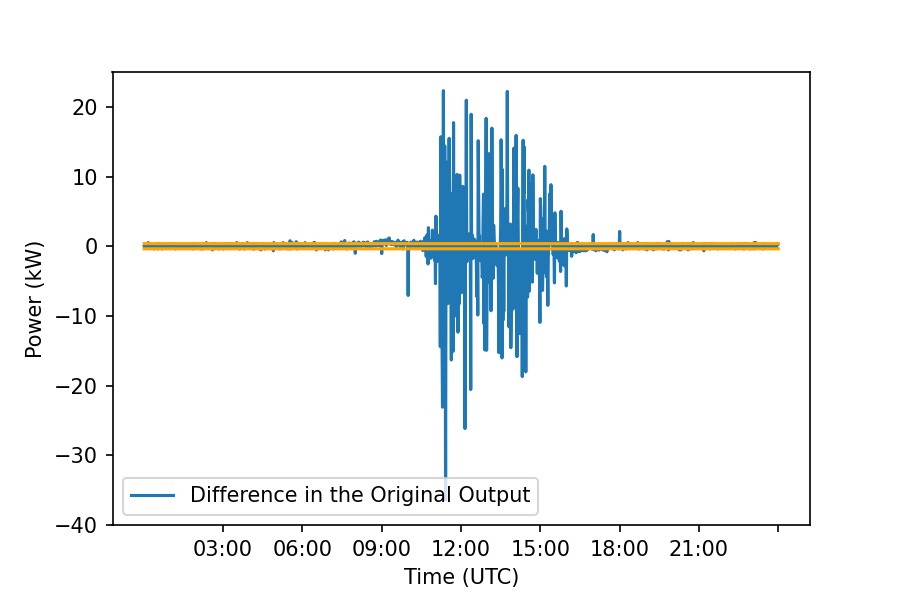}
\centering
\caption{The solar power generation difference compared to the ramp rate limit}
\label{Original_Difference}
\end{figure}

The goal of this paper is to control the battery charging and discharging rate as well as controlling the EV charging to maintain the changes in power output of the system within the ramp rate limit. Given the high penetration level of solar energy in a system, the changes in solar power will exceed the ramp rate capability of the system. The changes in power output are shown alongside the ramp rate limits of the system in the Nanogrid under study in Fig. \ref{Original_Difference}. The ramp rate limits are selected based on a 1\% change of maximum load within 1 minute. It is important to mention that the Nanogrid under study includes a very large PV array which is equal to 160\% of the maximum load installed.

\begin{figure}[h!]
\includegraphics[width=3.6in]{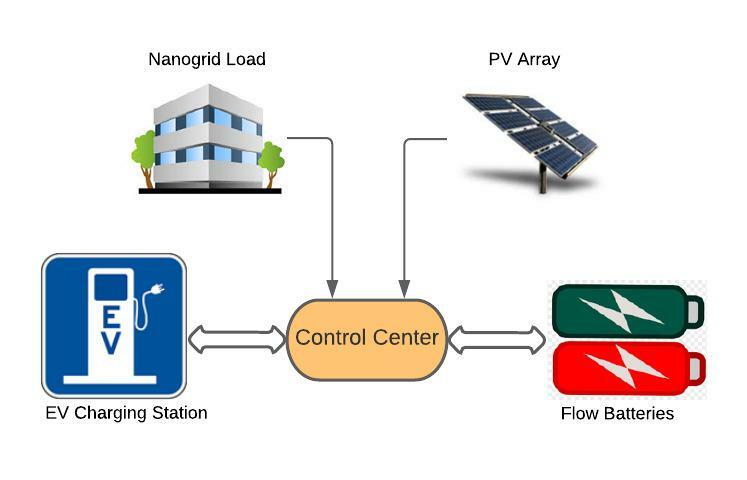}
\centering
\caption{The Nanogrid system under study}
\label{Nanogrid}
\end{figure}

The Nanogrid testbed includes PV generation, loads, flow batteries, and an EV charging station as illustrated in Fig. \ref{Nanogrid}. In this section we consider two different approaches to solve this problem. 

\subsection{Real-time Solar Smoothing}

The real-time control of the system will decide the charging rate of the battery and EV based on the current load and solar power data to minimize the violation of ramp rate limits. Therefore, if  the combination of PV output and load changes exceed the ramp rate limit, the battery and EV charging are assigned to make up the difference. The optimization formulation for the real-time solution is described in equations (\ref{OF}-\ref{EV_SoC}).

\begin{equation}
\begin{array}{rl}
\displaystyle \min & |(P_{PV}-P_{L}-P_{B}^{ch}+P_{B}^{Ddis}-P_{EV})-P_{ref}|\\
\label{OF}
\end{array}
\end{equation}
\begin{equation}
\begin{array}{rl}
\textrm{s.t.} & P_{B}^{Ch}<P_B^{max}\\
\label{BCH}
\end{array}
\end{equation}
\begin{equation}
\begin{array}{rl}
& P_{B}^{Dis}<P_B^{max}\\
\label{BDCH}
\end{array}
\end{equation}
\begin{equation}
\begin{array}{rl}
&  SoC_{B}^{min}\le SoC_{B}\le SoC_{B}^{max}  \\
\label{BSoC}
\end{array}
\end{equation}
\begin{equation}
\begin{array}{rl}
&  SoC_{EV}^{min}\le SoC_{EV}\le SoC_{EV}^{max}  \\
\label{EVSoC}
\end{array}
\end{equation}
\begin{equation}
\begin{array}{rl}
&  SoC_{B}=SoC_{B,t-\Delta t}+(P_{B}^{ch}\eta_{B}^{ch}-\frac{P_{B}^{dis}}\eta_{B}^{dis})\Delta t \times   \frac{100}{C_B} \\
\label{B_SoC}
\end{array}
\end{equation}
\begin{equation}
\begin{array}{rl}
&  SoC_{EV}=SoC_{EV,t-\Delta t}+P_{EV}\eta_{EV}^{ch}\Delta t  \\
\label{EV_SoC}
\end{array}
\end{equation}

The objective function which is defined in eq.(\ref{OF}) is set to minimize the violation from the ramp rate limits by selecting the appropriate charging rate for battery and EV. Where $P_{PV}$ is the PV output, $P_L$ is the system load, $P_B^{cH}$ and $P_B^{dis}$ are the charging and discharging rate of the battery and $P_{EV}$ is the EV charging rate. The target curve $P_{ref}$ in eq. (\ref{OF}) is equal to the summation of PV and load if no violation happens from the previous point. Otherwise, it will reflect the ramp rate limit. The ramp rate is the difference in overall system power considering PV output, load, battery, and EV charging. 

The limits for battery charging and discharging rates are considered in eq. (\ref{BCH}) and (\ref{BDCH}) where $P_B^{max}$ is the maximum rate of charging and discharging for the battery. 

The SoC constraints for battery and EV are described in eq. (\ref{BSoC}) and (\ref{EVSoC}) where $SoC_B^{min}$ and $SoC_B^{max}$ are the minimum and maximum SoC levels for the battery and $SoC_{EV}^{min}$ and $SoC_{EV}^{max}$ are the minimum and maximum SoC levels for the EV. It's important to mention the minimum SoC level for the EV is defined based on the current time, time of arrival, connection time and the available SoC level at the start of charging to ensure that the EV unit will be at least 90\% charged before leaving the charging station.

Lastly eq. (\ref{B_SoC}) and (\ref{EV_SoC}) apply the change in SoC levels of the battery and EV considering the charging rate where $SoC_{B,t-\Delta t}$ and $SoC_{EV,t-\Delta t}$ are the SoC levels of battery and EV at the time $(t-\Delta t)$, $\eta_B^{ch}$ and $\eta_B^{dis}$ are the efficiencies of the battery during charging and discharging, $C_B$ is the battery capacity in kWh, and $\eta_{EV}^{chg}$ is the EV charging efficiency. 

The real-time control is a single-step optimization. Any changes in the foreseeable future may be difficult to cope with because flexible resources may be used up.

\subsection{Predictive Solar Smoothing}

The predictive solar smoothing proposed in this paper aims to improve upon the real-time control by relying on solar power and load forecasts and preemptively deciding the charging rates of battery and EV to prepare for large changes in power output.

\subsubsection{Solar and Load Forecasting}

The forecasts of load and PV power are both generated using an artificial neural network with several layers. The number of neurons and layers in the neural network is decided based on the data used for training and validation to achieve the best performance in terms of accuracy. The input data to the neural network is the past instances of the PV power output and system load.

Once the forecasts are generated the next step is to shape a target curve for the overall system power. The goal of the target curve is to minimize or eliminate the ramp rate violations for the overall power output of the system while staying close to the original power output. Therefore, the objective function of the optimization in eq (\ref{OF}) will need to change in order to incorporate the target curve for all future times.

\begin{equation}
\begin{array}{ll}
\displaystyle \min & \\
\sum_{t}^{t+h}|(P_{PV,t}-P_{L,t}+P_{B,t}^{ch}+P_{B,t}^{dis}-P_{EV},t)-P_{ref,t}|\\
\label{MAOF}
\end{array}
\end{equation}

where we consider the overall violation in a future horizon $h$ and the target curve $P_{ref,t}$ is dependent on the solar power and load forecasts.

Therefore, the target curve in predictive solar smoothing not only considers ramp rate limits but also the future possible changes to solar power and load of the system. In this paper, we have two different approaches to shape the target power curve by looking at the forecasts.

\subsubsection{Moving Average target curve}

The target curve needs to consider the forecasts of changes in the future and smoothly raises or lowers the overall system power to operate within ramp rate limits while staying close to the original power output. The moving average target curve will stay close to the original power while smoothing any sudden changes in the power output profile.

\begin{equation}
\begin{array}{rl}
\displaystyle P_{ref,t}=\frac{1}{2n+1}\sum_{t-n}^{t+n}P_{PV,t}-P_{L,t}
\label{MA}
\end{array}
\end{equation}

\begin{figure}[h!]
\includegraphics[width=3.8in]{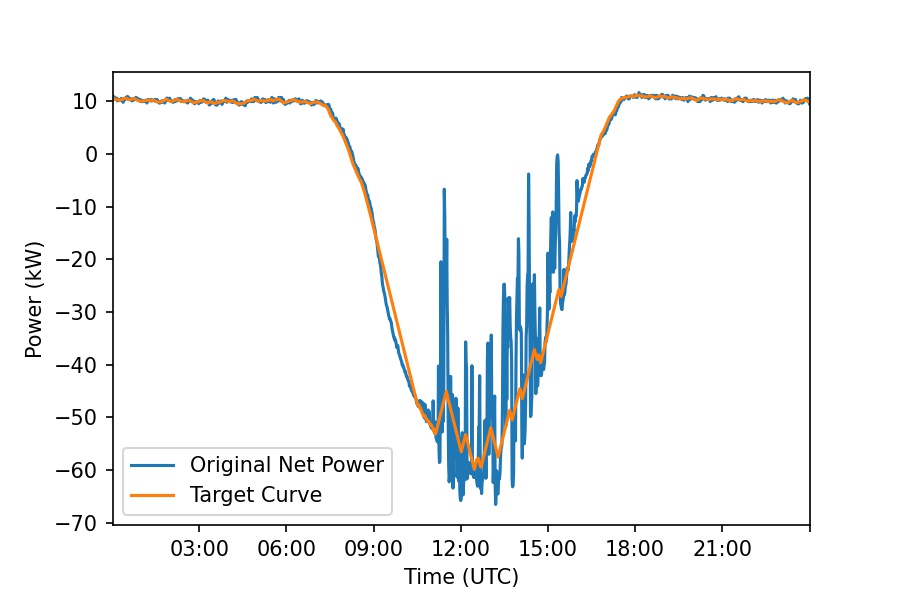}
\centering
\caption{The moving average target curve and original power of the system}
\label{target}
\end{figure}

This target curve will average the overall power output of the system from $n$ minutes ago to $n$ minutes into the future and generate a smooth curve compared to the original power output. Given the target curve which considers the forecasted power as well as previous power output, the control algorithm will be able to lower the violations from the ramp rate limits by preemptively lowering or raising the power output. Furthermore, the target curve is examined to ensure minimum ramp rate violations. The target curve compared to the original power output is shown in Fig. \ref{target}.

As visible in the figure, the target curve compared to the original system power is much smoother while staying as close as possible to the original power profile of the system. The target curve generated will not have any ramp rate violations but when applied to the control program the system might experience violations if the capacity of the battery and EV combined cannot compensate for the difference between the original power profile and the target curve. The changes in the target curve shown in Fig. \ref{target} will be limited to the ramp rate capability if there is enough battery and EV capacity available to follow the target curve.


The number of data points used in the moving average ($2n+1$) is selected based on a test that compares different moving average sizes and selects the one with the least ramp rate violations. It is important to consider that increasing the number of data points and the forecast horizon beyond a certain limit will decrease the accuracy of the forecasts and therefore diminish the benefits of the predictive model.

\subsubsection{Variance target curve}

Contrary to the moving average target curve, this target curve is primarily based on the variance of the forecasted power. The idea is to lower the net power of the Nanogrid when the forecast shows a high variance in the upcoming data which indicates changes to cloud coverage. Therefore, if the forecasted data has a high variance we will lower the power output to prepare for drops in solar power output due to cloud coverage or increase the output when the sky is clearing up. As a result, the battery and EV capacity is flexible to respond to solar power changes. The variance-based target curve for the power is calculated as below:

\begin{equation}
\resizebox{\columnwidth}{!}
{%
$\displaystyle P_{ref,t+h}=(1-\frac{\sqrt{Var_t^{t+h}P_{PV}}}{\sum_t^{t+h}P_{PV}})(P_{PV,t+h}-P_{PV,t})+P_{PV,t}-P_{L,t}
\label{VB}$%
}
\end{equation}

where $Var_t^{t+h}$ is the variance of solar power during the next h-minute window. After calculation, the ramp rate limits are applied to the target curve to avoid any possible ramp rate violations.

\section{Case Study}
\label{Case Study}

The proposed solar smoothing control algorithms in this study are applied to a Nanogrid system owned by OUC that contains PV generation, EV charging stations, and two units of flow batteries connected. The data for the PV output is gathered from the Nanogrid. The EV data is simulated considering historical data from other OUC sites. The load data is collected from the buildings connected to the Nanogrid via a power transformer. The Nanogrid capacity for PV and load has been scaled to represent a system with high penetration of solar power. The PV capacity installed is 80 kW, two flow battery units are connected with a capacity of 40kWh and a maximum charging rate of 10kW, four EV charging stations are connected to the Nanogrid and the maximum load installed is 40 kW.

The solar power and load forecasts are generated using the historical data from the OUC site. Additionally, the optimal configuration of the neural networks for both PV and load forecasting is selected based on tests performed on historical data. Furthermore, the number of data points used in the moving average target curve is selected based on historical data to optimize the target curve.



\begin{figure}[h!]
\includegraphics[width=3.8in]{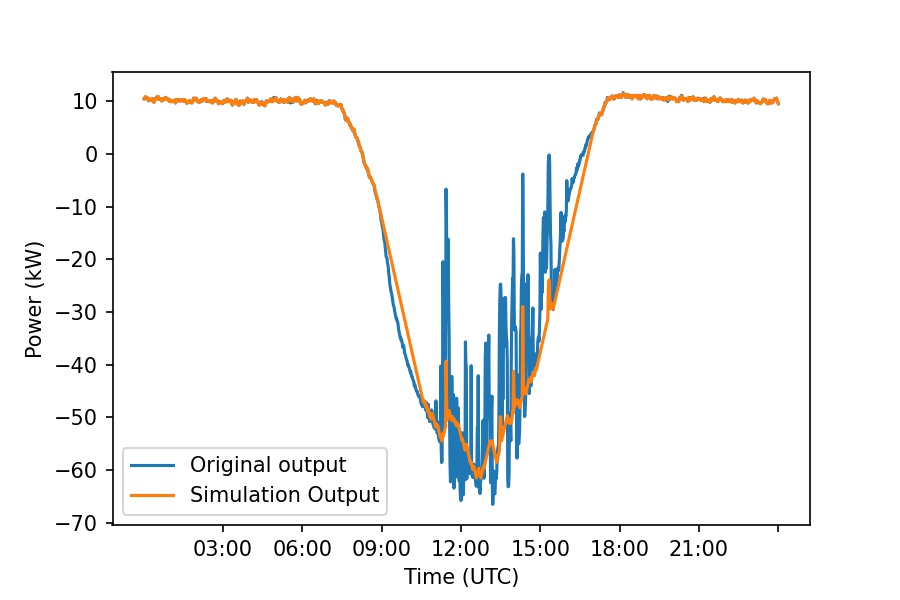}
\centering
\caption{The output of the real-time control compared to the original power output}
\label{real-time_output}
\end{figure}

The PV output and the load of the Nanogrid under study are combined and shown in Fig. \ref{real-time_output} as the original power output. The original net power of the Nanogrid is very volatile and exceeds the ramp rate limits by large margins as shown in Fig. \ref{Original_Difference}.


The first control algorithm applied for solar smoothing is the real-time control. This control method adjusts the charging rate of the battery and EV units based on real-time data to minimize ramp rate violations. The power output of this control method is shown in Fig. \ref{real-time_output} alongside the original output without the control. 

\begin{figure}[h!]
\includegraphics[width=3.8in]{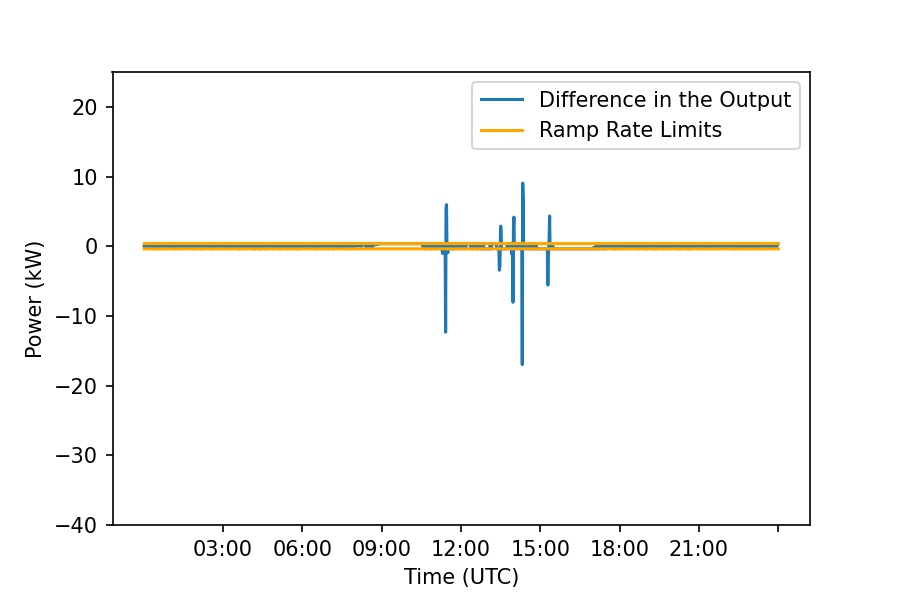}
\centering
\caption{The changes in power output compared to ramp rate limits for real-time control}
\label{real-time_diference}
\end{figure}

The real-time control is capable of maintaining the power output within the ramp rate limits for the most part but some violations exist that are shown in Fig. \ref{real-time_diference}. The figure shows the changes in the power output under the real-time control algorithm alongside the limits for ramp rate.

The predictive control compared to real-time control sets the charging rate of the battery and EV preemptively to prepare for the future changes in solar power output. Therefore, the results of the predictive control depend on the accuracy of the forecasts as well as the effectiveness of the control algorithm.

\begin{figure}[h!]
\includegraphics[width=3.8in]{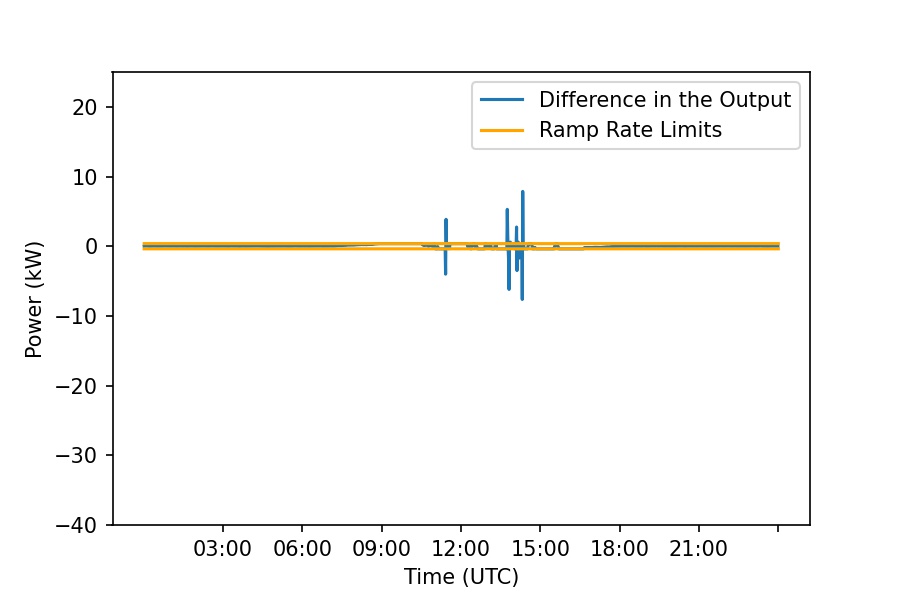}
\centering
\caption{The changes in power output compared to ramp rate limits for predictive control with moving average target curve}
\label{moving-average_diference}
\end{figure}

Next, we apply the predictive control algorithm with the moving average target curve to select the charging rates of battery and EV units based on historical data as well as forecasts. The goal of the predictive target curve is to set the charging rates for the battery and EV units to prepare for future changes in solar power output while considering the ramp rate limits in real-time. The power output changes for the predictive control with moving average target curve are shown in Fig. \ref{moving-average_diference}.

The figure shows an improvement in violations of ramp rate limits compared to the real-time control.

\begin{figure}[h!]
\includegraphics[width=3.8in]{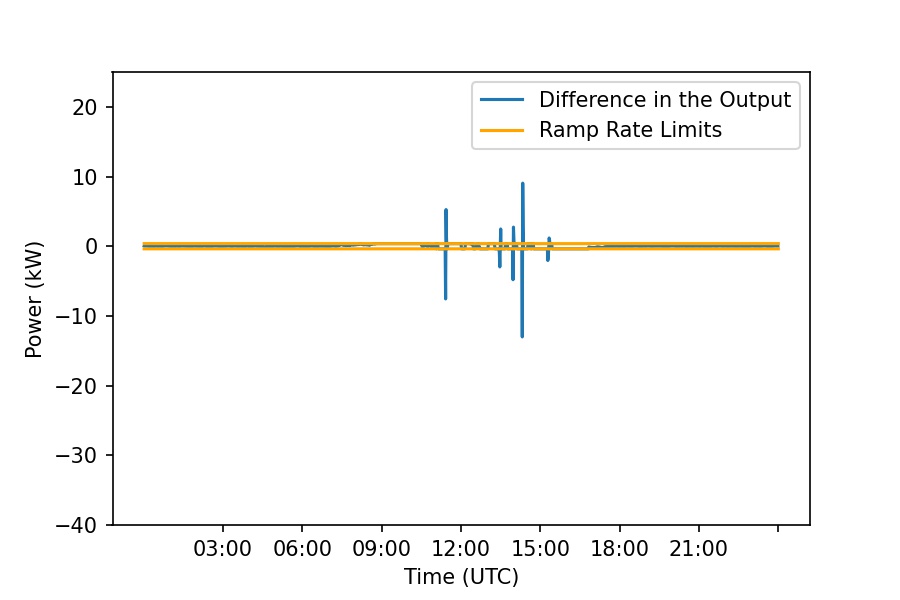}
\centering
\caption{The changes in power output compared to ramp rate limits for variance-based predictive control}
\label{variance_based_diference}
\end{figure}

The third control method applies the variance-based target curve to the optimization problem for the solar smoothing problem. While the variance-based model relies on forecasts similar to the moving average model it is not able to achieve the same level of improvement in terms of violations. Fig. \ref{variance_based_diference} shows the power changes under the variance-based predictive control. While the magnitude of violations from the ramp rate limit are smaller than the real-time control, it does not reduce the violations as much as the moving average method. The results of the control methods (sum of ramp rate violations) are shown in Table \ref{table}. The results show substantial improvement for the moving average model while the variance-based model also improves upon the real-time control model but not as much. 

\begin{table}[h]
\centering
\caption{Ramp Rate Violations for different control models}
\label{table}
\begin{tabular}{|l|c|}
\hline
Control Model & Sum of Violations (kW) \\ \hline
Real-Time & 240.99 \\ \hline
Predictive (Moving Average) & 113.56 \\ \hline
Predictive (variance-based) & 141.27 \\ \hline
\end{tabular}
\end{table}

\begin{figure}[h!]
\includegraphics[width=3.8in]{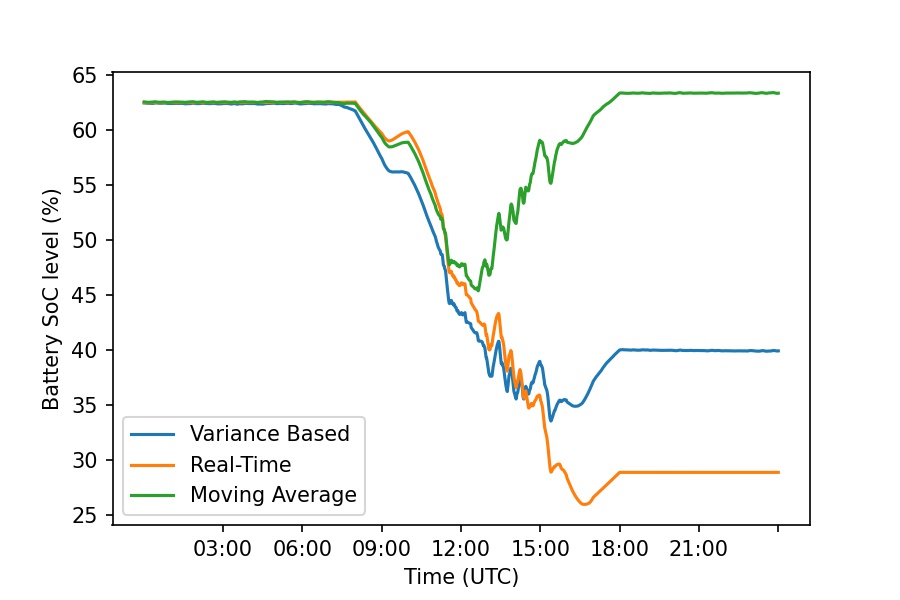}
\centering
\caption{The battery SoC level with different control methods}
\label{SoC}
\end{figure}

Another aspect of the control is the management of the resources such as the battery storage unit. In addition to better performance, the predictive model with the moving average target curve is able to manage the battery storage considerably better than the real-time control. As shown in Fig. \ref{SoC} the battery SoC level is much better recovered under the predictive control method. While the SoC level with the variance-based model improves the SoC level of the battery compared to real-time control, the moving average method is able to balance the charging and discharging the battery and recover the battery SoC level significantly towards the end of the day


The three control methods perform the same in terms of EV charging. The EVs are all fully charged at the end of the day and the charging profile looks similar for all three methods.

\section{Conclusion}

This study applies multiple solar smoothing control algorithms to a Nanogrid with PV, battery, and EV components to maintain the changes in the overall power profile to the ramp rate capability of the system. The effectiveness of the real-time control is shown in reducing the violations. Additionally, predictive solar smoothing control methods are proposed to preemptively control the battery and EV units and further decrease the ramp rate violations. The forecast of solar power generation and load of the Nanogrid are used to generate a target curve for the power profile which minimizes the violations while operating as close as possible to the original power profile. Furthermore, the proposed predictive control manages the battery storage unit much more effectively.

\label{Conclusion}

\bibliographystyle{IEEEtran}
\bibliography{ref}

\end{document}